\documentstyle[12pt,aps]{revtex}
  \font\elevenmib=cmmib10 scaled 1095
  \font\tenmib=cmmib10
  \font\eightmib=cmmib10 scaled 800
  \font\sixmib=cmmib10 scaled 667
  \skewchar\elevenmib='177
  \newfam\mibfam
  
  \textfont\mibfam=\tenmib
  \scriptfont\mibfam=\eightmib
  \scriptscriptfont\mibfam=\sixmib
  \mathchardef\alpha="710B
  \mathchardef\beta="710C
  \mathchardef\gamma="710D
  \mathchardef\delta="710E
  \mathchardef\epsilon="710F
  \mathchardef\zeta="7110
  \mathchardef\eta="7111
  \mathchardef\theta="7112
  \mathchardef\kappa="7114
  \mathchardef\lambda="7115
  \mathchardef\mu="7116
  \mathchardef\nu="7117
  \mathchardef\xi="7118
  \mathchardef\pi="7119
  \mathchardef\rho="711A
  \mathchardef\sigma="711B
  \mathchardef\tau="711C
  \mathchardef\phi="711E
  \mathchardef\chi="711F
  \mathchardef\psi="7120
  \mathchardef\omega="7121
  \mathchardef\varepsilon="7122
  \mathchardef\vartheta="7123
  \mathchardef\varrho="7125
  \mathchardef\varphi="7127

\def\sss#1{{\scriptscriptstyle #1}}

\def\ssr#1{{\sss{\rm #1}}}

\def\dsl{\raise.15ex\hbox{$/$}\kern-.57em\hbox{$\partial$}}
\def\nsl{\raise.15ex\hbox{$/$}\kern-.57em\hbox{$\nabla$}}
\def\id{\raise.72ex\hbox{$-$}\kern-.85em\hbox{$d$}\,}
\def\gtwid{\,{\raise.3ex\hbox{$>$\kern-.75em\lower1ex\hbox{$\sim$}}}\, 
}
\def\ltwid{\,{\raise.3ex\hbox{$<$\kern-.75em\lower1ex\hbox{$\sim$}}}\, 
}
\def\undr{\raise.3ex\hbox{$\sim$\kern-.75em\lower1ex\hbox{$|\vec  
x|\to\infty$}}}

\def\frac#1#2{{\textstyle{#1 \over #2}}}

\def\({\left (}
\def\){\right )}

\def\xhi{{\raise.35ex\hbox{$\chi$}}}

\def\undertext#1{$\underline{\hbox{#1}}$}

\def\and{a^{\phantom\dagger}}

\gdef\journal#1, #2, #3, 1#4#5#6{               % Journal reference.   
Comma sets
    {\sl #1~}{\bf #2}, #3 (1#4#5#6)}            % off: name, vol,  
\begin{document}
\draft
\def\uar{\uparrow}
\def\dar{\downarrow}
\def\mmz{{[m_\uar m_\dar 0]}}
\def\vhat{{\hat v}}
\def\rhobar{{\overline\rho}}
\def\sbar{{\overline s}}
\def\fifth{\frac{1}{5}}
\def\uu{{\uar\uar}}
\def\dd{{\dar\dar}}
\def\ud{{\uar\dar}}
\def\du{{\dar\uar}}
\def\izppi{\int\limits_{0^+}^{\infty}}
\def\IB{I_{\rm B}}
\def\gtw{{\tilde{g}}}
\def\ttw{{\tilde{t}}}
\def\thtw{{\tilde{\theta}}}
\def\tot{\frac{3}{2}}
\def\kF{{k_\ssr{F}}}
\def\kJ{{k_\ssr{J}}}
\def\thb{{\bar \theta}}
\def\phb{{\bar \phi}}
\def\Gb{{\overline G}}
\def\II#1#2{I_{#1}^{\ssr[#2\ssr]}}
\def\avg#1{{\overline{[#1]}}}
\def\Var{{\rm Var}}

%@twocolumnfalse\endcsname
\title{ Addition Spectrum Oscillations in Fractional Quantum Hall Dots}

\author{
Eyal Goldmann\\
Scot R. Renn }
\address{
Department of Physics, University of California at San Diego,
La Jolla, CA 92093}

\date{\today}

\maketitle

\begin{abstract}

Quantum dots in the fractional quantum Hall regime are studied using
a Hartree formulation of the composite fermion theory.  
Under appropriate conditions
the chemical potential of 
the dots 
will oscillate periodically with $B$ due to the
transfer of composite fermions between quasi-Landau bands.  This effect is analogous to
 the addition spectrum oscillations
which occur in quantum dots in the integer quantum Hall regime.
Period $\phi_0$ oscillations are found in 
sharply confined dots with filling factors $\nu = 2/5$ and $\nu = 2/3$.
Period $3\phi_0$ oscillations are found in a parabolically confined $\nu=2/5$
dot. More  generally, we argue  that the oscillation period of dots with band pinning should vary continuously with $B$ whereas the period of  dots without
 band pinning is $\phi_0$.
Finally, we discuss the possibility of 
detecting fractionally charged excitations using the observed period of
addition spectrum oscillations.

\end{abstract}


\vskip2pc

%\narrowtext

In the last few years there has been a great deal of interest in  
studying electron correlations and interaction effects in  quantum  
dots
and other highly confined geometries \cite{reviews,chakraborty}.
 One reason for this  interest  is the availability of experimental  
techniques such as single electron capacitance spectroscopy(SECS)
\cite{Ashoori,Ashoori2} and 
gated transport spectroscopy(GTS)\cite{McEuen,McEuen2} 
which allow one  to investigate modifications of the 
quantum dot addition spectrum  associated with these effects.
 Perhaps the most striking behavior exhibited 
 is the observed oscillatory field
 dependence\cite{Ashoori,Ashoori2,McEuen,McEuen2} 
 of the addition spectrum,
\begin{equation}
\mu_N(B)\equiv E_N(B)-E_{N-1}(B),
\end{equation}
which occurs when $\nu_0$,the filling factor 
at the droplet center, is approximately two. 
 This behavior results from the inter-Landau level transfer of 
electrons  which  occurs when the magnetic flux through the dot is  increased
 by approximately
$\phi_0=hc/e$.

Given this remarkable behavior, it is interesting to consider the 
possibility that related effects might occur in the fractional quantum
Hall regime.  The composite fermion (CF) picture of the FQH is particularly
suggestive, as it leads us to expect that
features similar to those at $\nu = 2$
might be observable at FQH filling factors for which the CF filling factor,
$\nu_{CF}$, is approximately two.
For example,  one would expect  oscillations similar to 
those seen in the $\nu_0=2$
dots\cite{Ashoori,McEuen} to occur for dots with $\nu = 2/5$ or $\nu = 2/3$.

%Ashoori et al.\cite{Ashoori} have reported   
%a series of bumps in the SECS 
%addition spectra which are attributed to   fractional quantum Hall
%effect physics. See Fig. 4 of ref. \cite{Ashoori}. 
%These bumps occur at filling factors 
%well below $\nu=2$ and exhibit a quasi-period of approximately $3\phi_0$.
%One possible interpretation of this behavior is  that the addition spectra
% oscillations are due to  the periodic transfer of 
%composite fermions(CFs) between
% quasi-Landau bands.

With this motivation, we have performed various Hartree calculations of
 the electronic 
structure of a parabolically confined dot with 
$\nu_0 = 2/5$ and rigidly confined 
dots with either $\nu=2/3$ or $\nu=2/5$.
The possibility of 
oscillations in the $\nu=2/3$ dot is rather interesting since their 
observation would provide indirect evidence for the
 $2/3\rightarrow 1 \rightarrow 0$ 
composite edge morphology proposed by Johnson
 and MacDonald\cite{compedge,Johnson,Heinonen}.

For the parabolically confined $\nu(0)=2/5$ dots, 
we indeed find  that $\mu_N(B)$ exhibits the expected
oscillations which occur 
with a period 
approximately given by $3\phi_0$. See Fig. 2.
%consistent with the observations by Ashoori\cite{Ashoori} .
In the cases of the sharply confined dots, oscillations also occur, but 
with periods of 
approximately $\phi_0$  in both the $\nu=2/5$ and $\nu=2/3$ cases.

Our approach to these calculations is based on a Hartree theory of 
composite fermions \cite{Brey,Chklovskii}.
According to CF theory,  the  fractional quantum Hall effect
is  a manifestation of the integer quantum Hall effect occurring in a  
weakly interacting gas of composite fermions\cite{Jain}.
This follows from the fact
that if  $\phi_p$ is the wavefunction  associated with  $p$ 
filled Landau levels,
 and $D=\Pi_{i < j}(z_i-z_j)^2$,  then the wavefunction of the form
$\chi =D^k \phi_p$ has
excellent overlap with  the exact $\nu=p/(2kp+1)$   ground state.
Alternatively, one may obtain the CF picture using 
the Chern-Simons(CS) construction,
 in  which one  attaches $2k$ fluxoids of fictitious  flux to each   
electron in the  2DEG\cite{Lopez,HLR}. According to this  approach, the  
CFs experience an effective magnetic field $\Delta  
B=B-\phi_0 2k\rho$. This implies an  effective filling  
factor $\nu_{\rm CF}\equiv n_{2D}/\phi_0 |\Delta B|=\nu/|1-2k\nu|$  
for the CFs.
It then follows that the  composite fermion liquid
with  $\nu_{\rm CF}=p$ is equivalent to an  incompressible 2DEG  with
\begin{equation}
\nu=p/(2kp\pm1),
\label{nu}\end{equation}
where the $+$ corresponds to $\Delta B > 0$ and the $-$ corresponds to 
$\Delta B < 0$.

There  is at least one unfortunate aspect of the Hartree   
approximation: It does not correctly give the mass of the CF.
In particular, the  Hartree approximation identifies the  
CF mass, $m_{\rm CF}$, with the electron band mass
($m_b=0.067m_e$ for GaAs). However, this is misleading since the fully
 renormalized $m_{\rm CF}$ 
 including all self-energy corrections 
must be independent of the band mass in the limit of no inter-Landau level
mixing.
 Moreover, at $\nu=1/2$, Chern-Simons gauge field corrections give rise to 
logarithmic divergent corrections to the band mass\cite{HLR}.  These 
corrections are believed to 
 be responsible for the apparent enhancement of $m_{\rm CF}$ occurs\cite{shubnikov}  as $\nu\rightarrow 1/2$.
  Since we do not wish to perform any
 sophisticated treatment of the CF  self-energy corrections, we will ignore this effect. 
Instead, we simply appeal to
 the dimensional analysis arguments of  Halperin, Lee, and Read\cite{HLR}. 
According to these authors,  in the absence of  Landau level mixing  the  
self-energy 
corrections  give a renormalized band mass 
$ m_{\rm CF} =m_0\sqrt{B}$ where $B$ is measured in Tesla and where $m_0$ 
is  $B$ independent.
To estimate $m_0$, we examined a variety of experiments regarding
 the composite fermion 
mass.These include the  activation gap measurements by 
R.R. Du et al.\cite{activationgap} 
which  give   $m_{\rm CF}=0.63m_{e}$ and $0.92m_e$ for electron 
densities of $1.12  
\times 10^{11} cm^{-2}$ and $2.3 \times 10^{11} cm^{-2}$,
respectively. We also considered the 
Shubnikov-de Haas oscillation experiments by Leadley\cite{shubnikov} 
which found  
$m_{\rm CF}=0.51+0.074|\Delta B|$ where 
$\Delta B=B-B_{1/2}$ and $B_{1/2}$ is the 
magnetic induction for $\nu=1/2$.
Based on these  results, we estimate $m_0$ to be roughly
$ 0.2m_e$.

The   Hamiltonian for $N$ composite fermions in a quantum dot is 
\begin{eqnarray}
{\rm \bf H}= \frac{1}{2m_{\rm CF}}& &\left(\sum^N_{i=1} (\vec{p}_i  
+\frac{e}{c}\vec{A}(\vec{r}_i) )-\frac{e}{c}\sum^N_{j\neq  
i}\vec{a}(\vec{r_i}-\vec{r_j})\right)^2 \cr
\ & \  & \qquad \qquad \qquad  +{\rm \bf U+V}
\end{eqnarray}
where 
\begin{equation}
\vec{a}(\vec{r_i}-\vec{r_j})=2k \frac{\hbar c}{e}  
\frac{\hat{z}\times(\vec{r}_i -\vec{r}_j)}{|\vec{r}_i-\vec{r}_j|^2}
\end{equation}
is the statistics potential, 
${\rm \bf U}=\frac{e^2}{2\epsilon}\sum^N_{i\neq j}, 
\frac{1}{|\vec{r}_i-\vec{r}_j|}$
is the Coulomb interaction, and ${\bf \rm  
V}=\sum^N_{i=1}V_C(\vec{r}_i)$ is the confinement energy,  where 
a parabolic 
confinement potential $V_C(r)=\frac{m_b \omega_0^2}{2}r^2$ is assumed.
In the calculations discussed below, we take $\hbar\omega_0 = 1.6$meV
and dielectric constant $\epsilon=13.6$, which is
 appropriate for the dots studied in ref. \cite{McEuen}.
Following Brey \cite{Brey} and Chklovskii \cite{Chklovskii}
we treat this problem using a Hartree  
approximation. The approximation involves
self-consistently solving the Poisson Schr$\ddot{\rm o}$dinger  
equation  
$H\phi_j(\vec{r})=\epsilon_j \phi_j(\vec{r})$
using the CF Hartree Hamiltonian\cite{Beenakker,Brey}
\begin{eqnarray}
H&=&\frac{1}{2m_{\rm CF}}(\vec{p}+\frac{e}{c}\vec{A}(\vec{r})  
-\frac{e}{c}\vec{A}_{CS}(r))^2\cr
\quad \ &  \  & \ \cr
 &+& \frac{e}{c}\int d\vec{r}' \  
\vec{a}(\vec{r}-\vec{r}')\cdot \vec{J}(\vec{r}') 
+V_H(\vec{r})+V_C(\vec{r})
\end{eqnarray}
where $V_H(\vec{r})$ is the Hartree potential,
$\vec{A}_{CS}(\vec{r})$ is the mean field statistics potential
\begin{equation}
\vec{A}_{CS}(\vec{r})= \int d^2\vec{r}' \vec{a}(\vec{r}-\vec{r}')
\rho(\vec{r}'),
\end{equation}

\begin{equation}
\rho(\vec{r})=\sum_{j=1}^N |\phi_j(\vec{r})|^2,
\end{equation}
and where
\begin{equation}
\vec{J}(\vec{r}) =\frac{1}{m_{\rm CF}} \sum_{j=1}^N
\phi^*_j(\vec{r})[\vec{p}+\frac{e}{c}\vec{A}(\vec{r})-\frac{e}{c}\vec{ A}_{CS}
(\vec{r})]\phi_j(\vec{r})
\end{equation}
is the charge current.

The chemical potential
of an $N$ particle droplet was taken to be the difference between the 
sums of single particle energies of dots with $N$ and $N-1$ electrons,
\begin{equation}
\mu_N(B)=\sum_{j=1}^{N} \epsilon_{j}(N,B) - 
\sum_{j=1}^{N-1} \epsilon_{j}(N-1,B).
\label{muN}\end{equation}
This method for computing $\mu_N(B)$ assumes that 
the energy required to remove the $k$th particle from the $N$-particle droplet 
may be identified with 
the single particle energy $\epsilon_k(N,B)$.  For Hartree-Fock
Hamiltonians, Koopman's theorem says that this identification is exact 
\cite{March}
provided that one neglects the effect of particle removal on the remaining
eigenstates.  However, because of the  three-body interactions in
 the CF kinetic energy, Koopman's theorem is not strictly applicable. 
Nevertheless,
for the purposes of identifying
addition spectrum oscillations, eq. \ref{muN} should be adequate.
 
{\it Numerical Results.}
The first systems  studied were  parabolically confined dots with 58--60
 electrons and $\nu(0) \approx 2/5$.
In Fig. 1(a), we present a density profile of an $N=60$ dot at 
$B=8.1 \rm T$.
The  shape of this density profile approximates the electrostatic 
solution of a parabolically confined classical 2DEG\cite{Shikin},
\begin{equation}
\rho_{es}(r) = \rho_0 (1-r^2/R^2)^{1 \over 2}
\label{essoln}\end{equation}
where $R^3= {3\pi \over 4} {e^2 N \over \epsilon m_b \omega_0^2}$ and 
$\rho_0 = {3N \over 2\pi R^2}$.
However, a closer inspection of Fig. 1(a) reveals 
a series of plateau-like features 
occurring at FQHE filling factors\cite{Glazmanetal}.
These features include a robust $\nu=2/5$ droplet at the center,
and  a 
$\nu \approx 1/3$ shoulder between $90{\rm nm} < r <130 {\rm nm}$.
To determine the $B$ dependence of $\mu_N(B)$, the Hartree equations were
self-consistently solved
over a range 
of $B$ for droplets with $N = 58$--$60$.
The results are plotted in Fig. 2. 
In this figure, we have labeled the   maxima of  $\mu_N(B)$ with $(n_1,n_2)$
 where  $n_i$ is  the  occupation of the $i$th quasi-Landau band (QLB).
The behavior observed is quite analogous to the results for the IQH regime dots:
With increasing $B$, $\mu_N(B)$ exhibits oscillations associated 
with the periodic
 transfer of composite fermions from the second 
quasi-Landau level into the first.
There is, however, one significant difference 
between the IQH and FQH calculations
 viz. the observed  periodicity. We defined  
$\Phi$ to be the flux inside a disk of
 radius $R=185$nm, as indicated with the arrow in  Fig. 1(a). 
 The  oscillation
 period $\Delta \Phi$ is then found to be 
$3.1\pm0.3\phi_0$. 
%Approximately  three fluxoids are required
% in order to transfer a composite fermion from the interior to the 
% boundary of the dot.

We have also studied dots which are confined rigidly.  In these 
calculations, the confinement potential is provided by a circular disk of
uniform positive charge $N_+e$ and charge density $\rho_+$ which terminates
in an infinite potential barrier of radius $R$.  For the studies of rigidly 
confined $\nu=2/5$ dots we took $N_+=60$ and $N=55$ or $56$, with $R$ set so 
that $\nu_+={\phi_0 \rho_+ \over B}=2/5$ at $5 {\rm T}$.
In Fig. 1(b), we present the  density 
profile of an $N = 56$ dot at $B=8.55T$.  
Observe that the density profile of the rigidly confined 
dot is quite different from that 
of the parabolically confined dot.  Whereas the density profile of the 
latter is hemispheric, that of the former is flat in the interior, 
except for a mound between $30 {\rm nm} < r < 120 {\rm nm} $ due to the
presence of states in the second QLB.
As in the case of the parabolically confined dot, $\mu_N(B)$ for this system 
oscillates due to interband transfers of CFs.  In this case, we find period
$1.03 \pm 0.15 \phi_0$.

The case of dots with $\nu=2/3$ in the interior is of particular interest 
since exact diagonalization studies by 
 Johnson and MacDonald\cite{Johnson} indicate that the local filling factor
 may exhibit an unusual $2/3 \rightarrow 1 \rightarrow  0$
 morphology at the edge.
%Both electroneutral and non-electroneutral dots will be 
%considered, so let  $N_+$ be the total neutralizing charge
% and let $N$ be the number of electrons.
For these studies,  we took $N_+=100$, setting $R$ so that 
$\nu_+ = 2/3$ at $5.0$ T.
In  Fig. 1(c), density profiles of
 a $(N,N_+)=(101,100)$ and a $(110,100)$ dot are presented.
In these plots, $\nu(r)$ is quite flat in the dot interior, but rises sharply
near the edge, and then drops to zero at the barrier.  In the case of the 
$N=110$ dot, $\nu(r)$ peaks at $\nu \approx .96$, a consequence 
of the occupation of only one QLB near the droplet edge.
The observed period for the chemical potential oscillations of these dots is 
$1.2 \pm 0.1 \phi_0$ for the $N=101$ dot and $1.03 \pm .15 \phi_0$
for the $N=110$ dot.  Parabolically confined $\nu_0=2/3$ dots were
not considered because in such a system
$\rho(r)$ would pass through $\nu=1/2$ gradually, thus forcing the 
occupation of additional QLBs and complicating the 
inter-band transfers.

{\it Discussion:}  
The $\phi_0$ periods of the rigidly confined dots may be understood as follows:
First recall that a state in the first quasi-Landau level with angular momentum
 $l$ lies on the equipotential which encloses  $l$ of effective (external minus
 fictitious) flux quanta.  For
 dots in which   the first quasi-Landau level is unpinned, the angular momentum of the outermost orbit $l_{max}$ coincides with $n_1$, the occupancy of the lowest qLL. Hence
\begin{equation}
n_1\phi_0 \approx \Phi_{ext}(n_1;N)-2(N-1)\phi_0
\end{equation}
where the first and second terms 
on the right-hand side are, respectively, the external flux and the fictitious
flux.  The above equation implies that if an additional unit of external 
magnetic flux is introduced, then $n_1$ increases by one i.e. an inter qLL 
transfer of a composite fermion will occur. Therefore the period of addition
 spectrum oscillations is unity for dots in which the first quasi-Landau level
 is unpinned. It should be noted that the quantisation of the oscillation
 period is precise only in the semiclassical limit. This follows from the 
fact that the radius of the droplet has an uncertainty of order the magnetic length. This implies that the precision of the oscillation period is at most $2l'_m\phi_0/R$ where $l'_m=(\hbar c/eB_{eff})^{1/2}$ is the magnetic length of the composite fermions.

%a simple argument, which we present for the case of the $\nu(0)= 2/5$
%(the case of the $\nu(0) = 2/3$ dot is similar).  We assume that 
%the dot phase separates into a well defined $\nu = 1/3$ region and a well 
%defined $\nu=2/5$ region.  For band occupancies $(n_1,n_2)$, the $\nu=1/3$
%and $\nu=2/5$ regions are habitated by $(n_1-n_2)$ and $2n_2$ composite
%fermions, respectively.  The total number of fluxoids penetrating the dot 
%is therefore
%\begin{equation}
%\Phi(n_1,n_2) = \phi_0[3(n_1-n_2)+{5 \over 2}2n_2] = \phi_0(3n_1+2n_2).
%\end{equation}
%Consequently, the additional flux required for a band transfer 
%$(n_1,n_2) \rightarrow (n_1+1,n_2-1)$ is $\phi_0$.

The above argument for period $\phi_0$ oscillations fails for the parabolically confined dot because 
the first QLL is pinned at the Fermi level.
In that case, the estimate described below is more appropriate. First, we note that in the semiclassical limit, 
\begin{equation}
\frac{N_2}{2k+1}=\int d^2 r \ \ (\rho(r)-\rho_L)\theta(\rho(r)-\rho_L)
\label{Ntwo}\end{equation}
where $\rho_L=\frac{1}{|2k+1|}B/\phi_0$ and where $\theta(x)=1$ for $x>0$ 
and $\theta(x)=0$ otherwise. This result is obtained as  follows. The charge 
density  $\rho(r)=\rho_1(r)+\rho_2(r)$ where $\rho_i(r)$ is the density 
associated with composite fermions in the i-th quasi-Landau band. Now in region
 of the dot where $\rho_2(r)>0$,  the lowest quasi-Landau band is fully 
occupied. This means that $\rho_1(r)=B_{eff}(r)/\phi_0$  
where  $B_{eff}(r)=B-2k \phi_0 \rho(r)$.
One then readily obtains the equation $N_2=\int d^2 r \  (\rho(r)-B_{eff}(r)/\phi_0)\theta(\rho_2(r))$ which, in turn, gives eq. \ref{Ntwo}.  Now the density profile of the parabolic dot is accurately given by the the elecrostatic profile $\rho_{es}(r)$. We insert this into our expression for $N_2$, eq. \ref{Ntwo}, and then we  calculate the period using $\Delta \Phi=\pi R^2[B(N_2)-B(N_2+1)]$. In this manner, we find
\begin{equation}
\frac{\Delta \Phi}{\phi_0} \approx \frac{(3\nu_0)^2}{(3\nu_0)^2-1}
\end{equation}
where $\nu_0=n_0 \phi_0/B$.
For $8.4 > B > 7.8$, the filling fraction of the  $60$ electron 
classical (electrostatic) dot is
$0.40 < \nu_0 < 0.43$. Hence the oscillation period of the semiclassical dot 
lies in the interval $2.5 < \Delta \Phi/\phi_0 < 3.3$.
This is consistent with the observed $\Delta \Phi/\phi_0 =3.1\pm 0.3$ period obtained from 
the Hartree calculation. One can obtain\cite{unpub} a  more precise estimate ($3.0<\Delta \Phi/\phi_0<3.3$) using eq. \ref{Ntwo} together with a density profile
 which includes a $\nu=2/5$ core as occurs in the dot illustrated in  Fig. 1 a.

Several additional observations are worth making: The period $3\phi_0$ 
oscillations of the parabolically confined $2/5$ dot are only indirectly
 associated with the existence of charge $e/3$ excitations.
To understand the connection between the oscillation period and charge
 fractionalization, we will modify eq. \ref{Ntwo} so as to allow  only 
charge $e$ transfers. This is done by replacing the left hand side of
 eq. \ref{Ntwo} with $N_0$ where
 $N_0$ is the integral number of electrons inside the incompressible strip. 
A quick inspection of the modified eq. \ref{Ntwo}, reveals that, $\Delta \Phi_e$, the period 
allowing charge $e$ transfer, is related to $\Delta \Phi$, the  period associated with fractional charge transfer according to
 $\Delta \Phi_{e}=\Delta \Phi/q$ where $q=1/(2k+1)$ is the quasi-particle
 charge or equivalently the local 
charge of the composite Fermion. Hence, in absence of charge fractionalization,  the  period of the dots
 shown in fig. 1 a-c would approximately be $9\phi_0$, 
$3\phi_0$, and $\phi_0$ respectively. Hence, we conclude that an experimental
 observation of period $\phi_0$ 
oscillations for a $2/5$ dot would be evidence for fractionally charged
 excitations. The experimental observation of charge $3\phi_0$ oscillations would be consistent with a hemispherical dot and $e/3$ charge fractionalization.
 However,  such a period would  also be consistent with charge $e$ excitations
 in a dot without band pinning.  

In summary, we have performed a series of Hartree calculations which
demonstrate the existence of addition spectrum oscillations associated with 
the transfer of composite fermions between quasi-Landau levels. The 
period of these oscillations depends on the presence or absence of band
pinning which, in turn, depends on the nature of the confinement potential.
In the absence of band pinning one observes period $\phi_0$ oscillations.
But in dots with band pinning,  spectral oscillations with unquantized periods occur. In some systems, the observed oscillation period might be used
to detect fractionally charged excitations.

{\it Acknowledgements:} S.R. would like to acknowledge support from the
 Alfred P. Sloan foundation, from NSF grant DMR 91-13631, and from the
 Hellman foundation. We would also like to acknowledge useful
 conversations and communications with D.P. Arovas, 
R.C. Ashoori, D.B. Chklovskii, and
 M.D. Johnson.

\vspace{0.4 in}

\undertext{FIGURE CAPTIONS}

\vspace{0.1 in}

FIG.1. (a) Density profile  for a quantum dot consisting of 60 electrons.
The center of the dot is a $\nu=2/5$ droplet.  The dotted line indicates
the density profile predicted by classical electrostatics.
(b) Density profile for a sharply confined dot of N=$56$ electrons, 
background charge $60e$, and $\nu \approx 2/5$.
(c) Density profiles  
for two sharply confined quantum dots with $N=101$ and
 $110$ electrons, and $\nu \approx 2/3$ in the bulk.
The background charge is $100e$ for both dots. 

FIG. 2 Chemical potential oscillations for 
parabolically confined quantum dots with $\nu=2/5$ and $59$ or $60$ 
electrons. At a peak labeled by $(n_1,n_2)$,  
there are $n_1$ composite fermions in the 
first quasi-Landau band and $n_2$ composite fermions in the second 
quasi-Landau band.
The oscillation  period is approximately $3\phi_0$.

FIG. 3 Chemical potential oscillations for a rigidly confined quantum dot
containing electronic charge $56e$ and background charge $60e$ near $\nu=2/5$.
The oscillation period is approximately $\phi_0$.
The $(n_1, n_2)$ notation
is explained in fig. 2.

FIG. 4 Chemical potential for sharply confined quantum dots with 
$\nu \approx 2/3$ 
as a function of magnetic field.  The solid and dashed  curves show
$\mu(B)$ for  dots with 101 and 110 electrons, respectively.  Both  dots 
have a background charge $100e$. The oscillation periods are
approximately $\phi_0$, and 
expanded views of the oscillatory behavior 
are provided in the insets. The $(n_1, n_2)$ notation 
is explained in fig. 2.


\begin{references}



\bibitem{reviews} T. Chakraborty, Indian Jour. of Pure and Appl. Phys. 
{\bf 32}, 575 (1994).
D. Heitmann and J.P. Kotthaus, Phys. Today, 56 (1993).

\bibitem{chakraborty}T. Chakraborty Comm. Cond. Mat. Phys. {\bf
16}, 35 (1992).

\bibitem{Ashoori} R.C. Ashoori, H.L. Stormer, J.S. Weiner, L.N.  
Pfeiffer, K.W. Baldwin, and K.W. West, Surf. Sci. {\bf 305}, 558 (1994). 

\bibitem{Ashoori2} R.C. Ashoori, H.L. Stormer, J.S. Weiner, L.N.
Pfeiffer, K.W. Baldwin, and K.W. West, Phys. Rev. Lett. {\bf 71}, 613  
(1993); Physica B {\bf 189}, 117 (1993); Phys. Rev. Lett. {\bf 68},  
3088 (1992).

\bibitem{McEuen}P.L. McEuen et al., Phys. Rev. Lett. {\bf 66} 1926  
(1991); M.A. Kastner, Rev. Mod. Phys. {\bf 64}, 849 (1992). 

\bibitem{McEuen2}
P.L. McEuen, E.B. Foxman, Jari Kinaret, U. Meirav, and M.A. Kastner, Ned  
Wingreen, S.J. Wind, Phys. Rev. B {\bf 45}, 11419 (1992).

\bibitem{compedge} A.H. MacDonald, Phys. Rev. Lett. {\bf 64}, 220 (1990).
\bibitem{Johnson}  M.D. Johnson, A.H. MacDonald, Phys. Rev. Lett.  
{\bf 67}, 2060 (1991).

\bibitem{Heinonen} O. Heinonen, M.I. Lubin, and M.D. Johnson,
Phys. Rev. Lett. {\bf 75}, 4110 (1995)

%\bibitem{jerks} Hereafter, we will refer to the  (101,100) and (110,100) dots
%as the electroneutral and the  non-electroneutral dots.

\bibitem{Brey}  Luis Brey, Phys. Rev. B{\bf \ 50, }11861 (1994).

\bibitem{Chklovskii} D.B. Chklovskii, Phys. Rev. B{\bf \ 51, }9895 (1995)

\bibitem{Jain} J.K. Jain, Phys. Rev. Lett. {\bf 63}, 199 (1989); Adv.  
Phys. {\bf 41}, 105 (1992).

\bibitem{Lopez}A. Lopez and E. Fradkin, Phys. Rev. B {\bf 44}, 5246  
(1991); see also E. Fradkin, {\it Field Theories of Condensed Matter  
Physics}(Addison Wesley, MA, 1991).

\bibitem{HLR}B.I. Halperin, P.A. Lee, and N. Read, Phys. Rev. B {\bf  
47}, 7312 (1993).

\bibitem{shubnikov}D.R. Leadley, R.J. Nicholas, C.T. Foxon, and J.J.  
Harris, Phys. Rev. Lett. {\bf 72}, 1906 (1994);
R.R. Du, H.L. Stormer, D.C. Tsui, L.N. Pfeiffer, and K.W. West, Solid  
State Comm. {\bf 90}, 71 (1994).

\bibitem{activationgap} R.R. Du, H.L. Stormer, L.N. Pfeiffer, K.W.  
Baldwin, and K.W. West, Phys. Rev. Lett. {\bf 71}, 3850 (1993).

\bibitem{Beenakker}  C.W.J. Beenakker and B. Rejaei, Physica B {\bf  
189, } 147 (1993).

\bibitem{March} See, for example, W. Jones and N. March, {\it Theoretical Solid
State Physics Vol. 1} (Wiley-Interscience, New York, 1973).
\bibitem{Shikin} V. Shikin, S. Nazin, D. Heitmann, and T. Demel, Phys. Rev.
B {\bf 43}, 11903 (1991).
\bibitem{Glazmanetal}  D.B. Chklovskii, B.I. Shkovskii and L.I.  
Glazman, Phys. Rev. B {\bf 46}, 4026 (1992);  D.B. Chklovskii and P.A. Lee, 
Phys. Rev. B {\bf 48}, 18060 (1993).


\bibitem{unpub} E. Goldmann and S.R. Renn, unpublished.

\end{references}
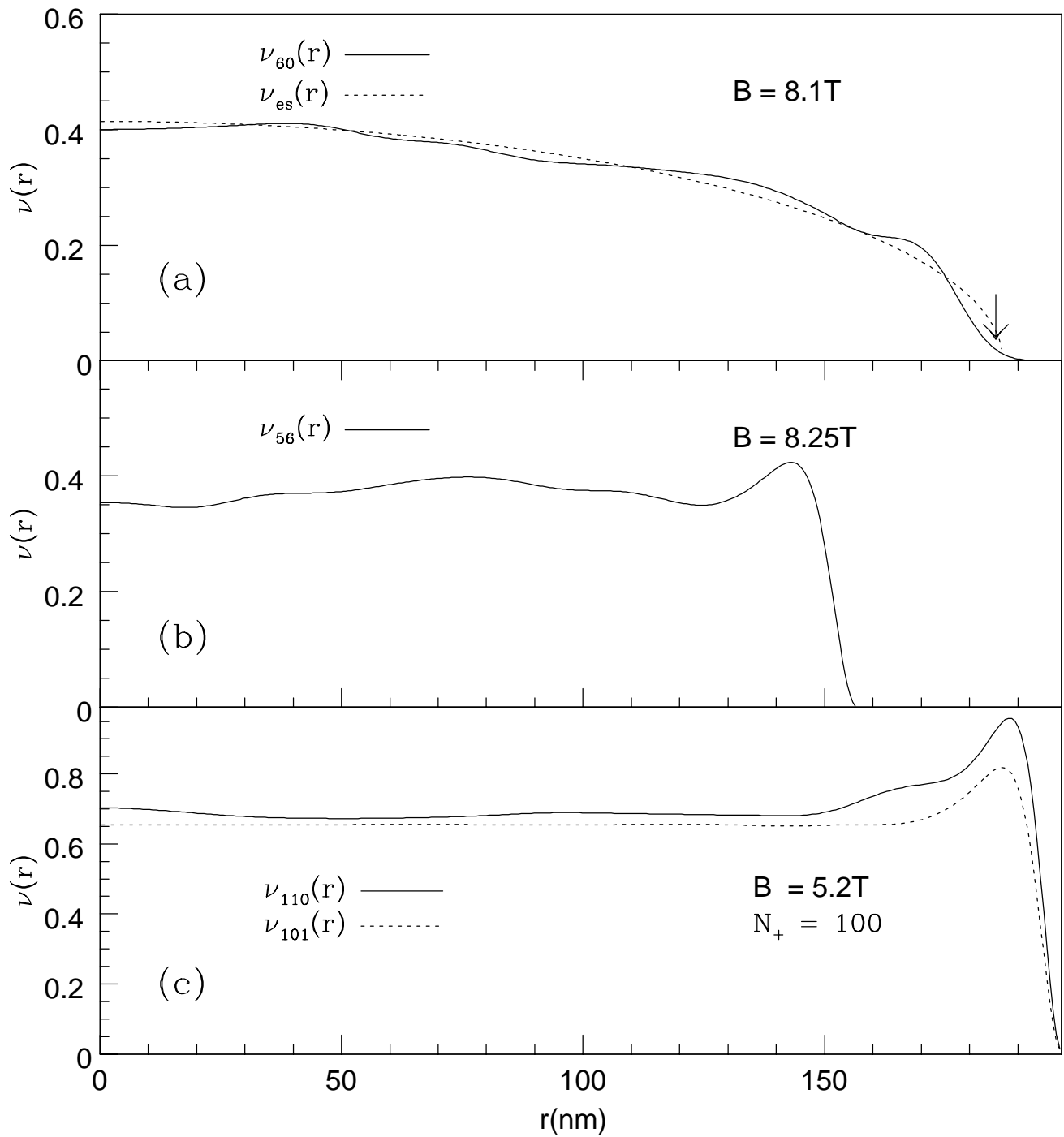
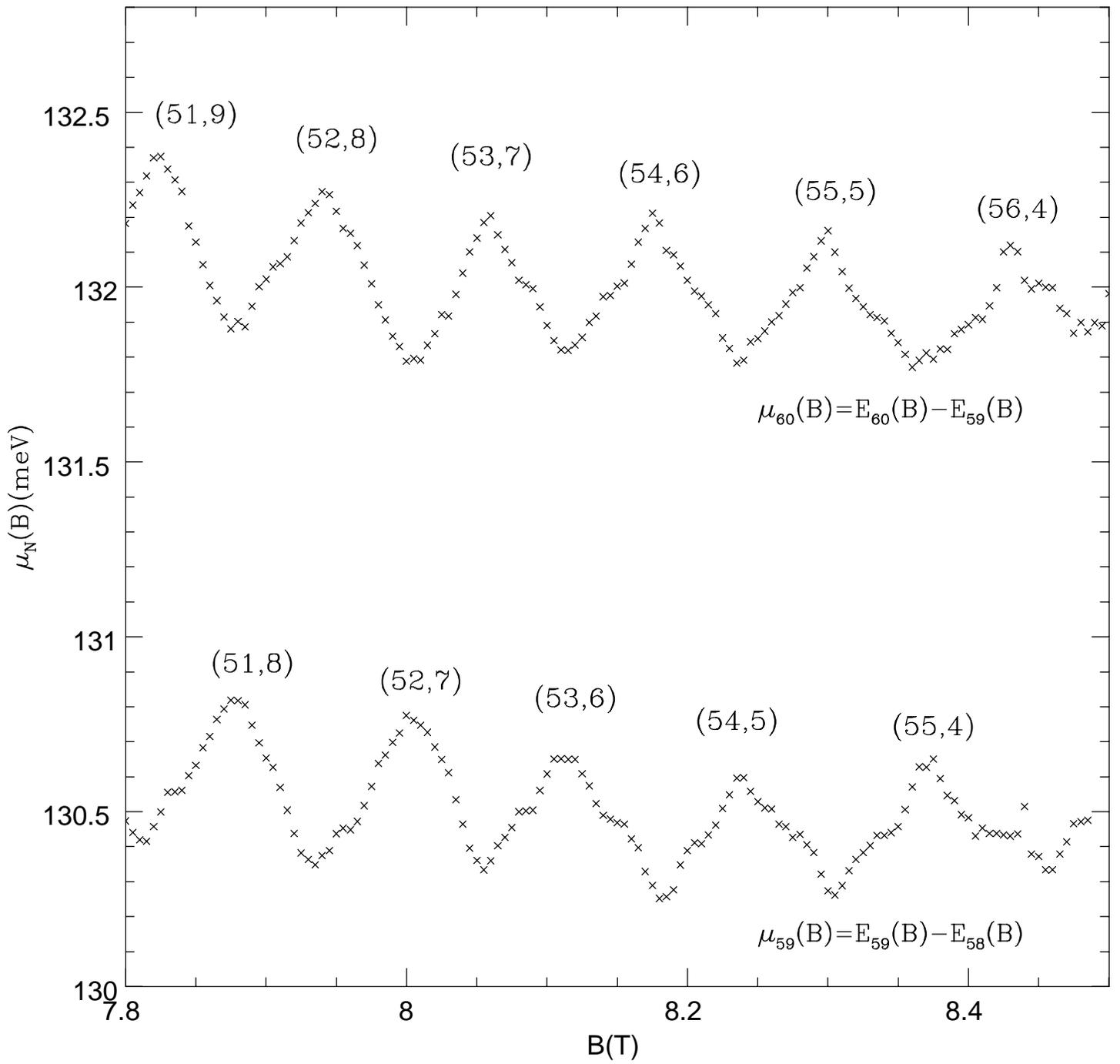
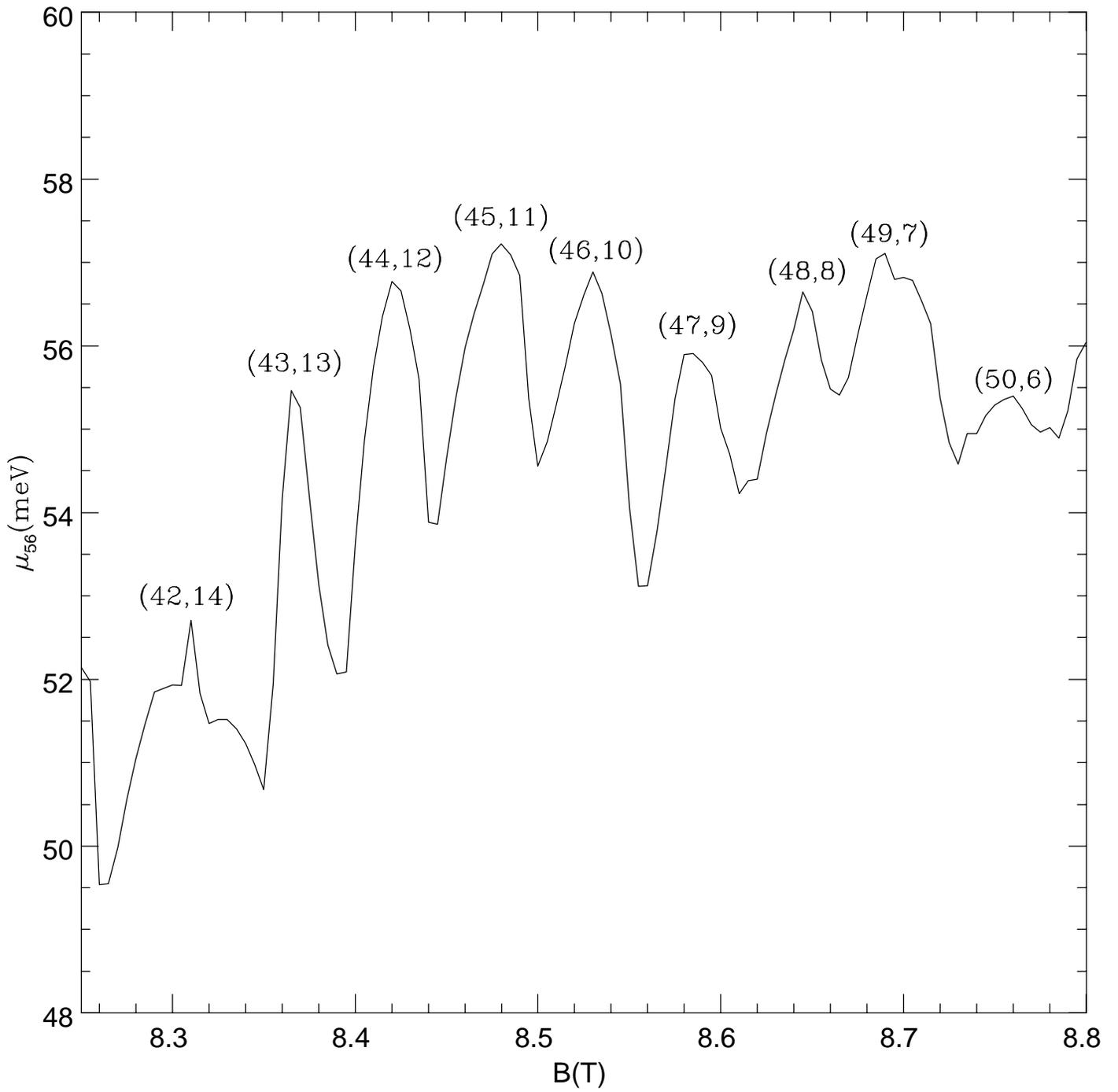
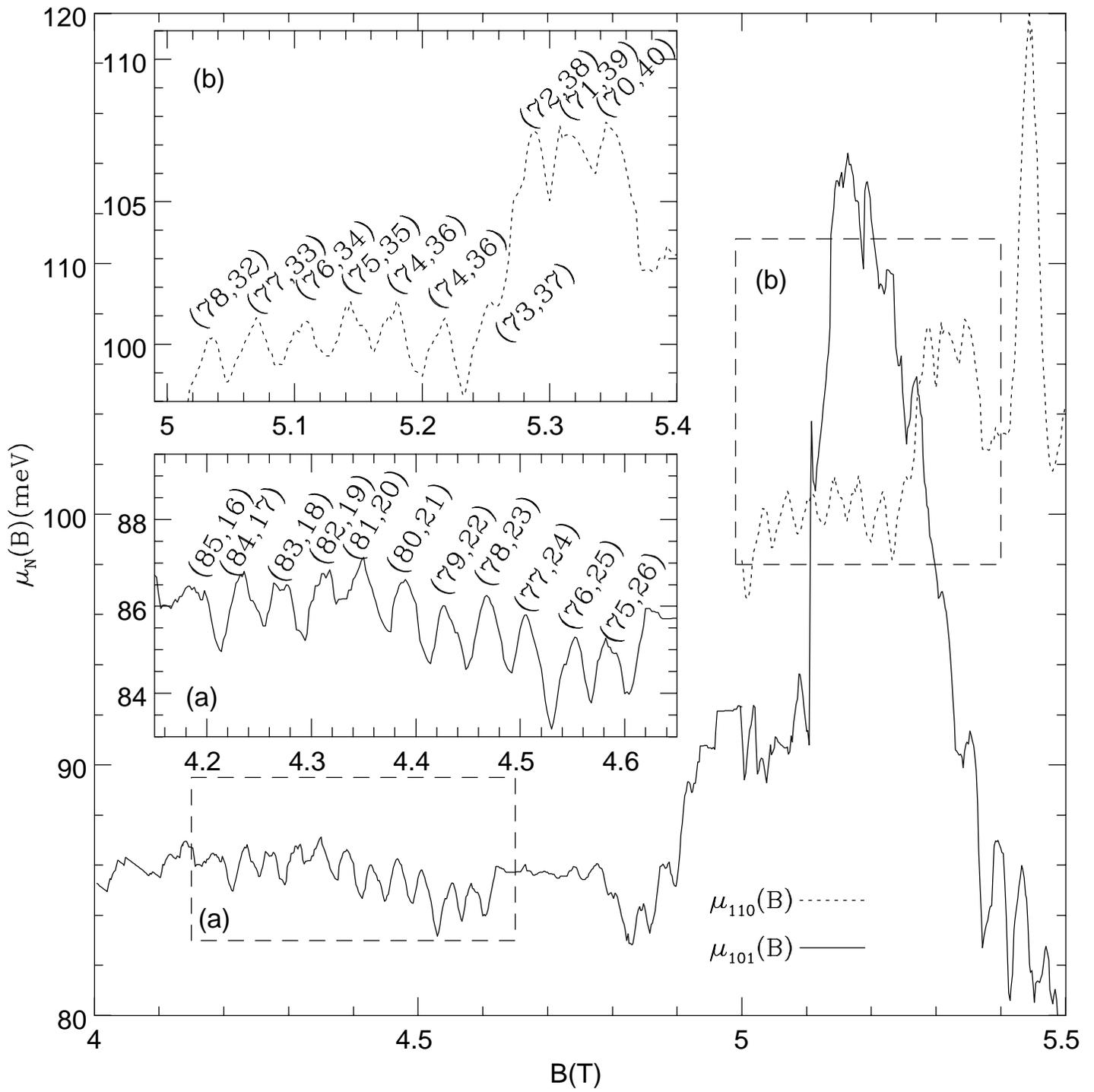
\end{document}